# Analysing the spotless days as predictors of the solar activity from the new sunspot number

V.M.S. Carrasco[1,2] • J.M. Vaquero[3,4] • M.C. Gallego[2,3]

[1] Centros de Estudios Energéticos y Radiofísicos, Badajoz, Spain

[2] Departamento de Física, Universidad de Extremadura, Badajoz, Spain

[3] Departamento de Física, Universidad de Extremadura, Mérida (Badajoz), Spain

[4] Instituto Universitario de Investigación del Agua, Cambio Climático y Sostenibilidad (IACYS), Universidad de Extremadura, Badajoz, Spain

**Abstract.** The use of the spotless days to predict the future solar activity is here revised based on the new version of the sunspot number index with a 24-month filter. Data from Solar Cycle (SC) 10 are considered because from this solar cycle the temporal coverage of the records is 100 %. The interrelationships of the timing characteristics of spotless days and their comparison with sunspot cycle parameters are explored, in some cases finding very strong correlations. Such is the case for the relationship between the minimum time between spotless days either side of a given solar maximum and the maximum time between spotless days either side in the previous solar minimum, with $r$ = -0.91 and a $p$-value < 0.001. However, the predictions for SC 24 or 23 made by other authors in previous works using the spotless days as a predictor of solar activity are not good since it has not been fulfilled. Although there seems to be a pattern of strong correlation for some relationships between the parameters studied, prediction of future solar cycles from these parameters defined as functions of the spotless days should be made with caution because sometimes the estimated values are far from the observed ones. Finally, SC 23 seems to show a mode change, a break respect to the behavior of their previous solar cycles and more similar to SC 10-15.

**1. Introduction**

Telescopic sunspot records span the period from 1610, with the first observations made by Harriot, Scheiner or Galileo, to the present, forming the most extensive dataset related to direct observations of solar activity (Vaquero and Vázquez, 2009). For this reason, the family of sunspot number indices is one of the most commonly employed in works about long-term solar activity. The two main indices used to describe the behaviour of solar activity from sunspot observations have been the international sunspot number and the group sunspot number (Hoyt and Schatten, 1998; Clette *et al.*, 2007). However, these indices present important differences in some periods, especially in the historical part (Clette *et al.*, 2014). For this reason, a recent revised collection of the sunspot group numbers has been made by Vaquero *et al*. (2016), and several new indices from sunspot observations have been presented (Clette *et al*., 2015; Usoskin *et al*., 2016; Svalgaard and Schatten, 2016).

The study of solar activity is especially interesting in several scientific fields, for example, climate change or the solar dynamo (Haigh, 2007; Charbonneau, 2010). Moreover, several phenomena of strategic importance for society, such as the radiative budget in our atmosphere or space weather, are related to future solar activity. Therefore, its prediction is important due to its impact on the human life (Pulkkinen, 2007; Pesnell, 2012). Petrovay (2010) pointed out that prediction methods can be divided into three groups: (i) precursor methods that rely on the previous values of the solar activity parameters to predict the following solar cycles; (ii) extrapolation methods which assume that the physics which generates the sunspot number record is statistically homogeneous at any point of time, and therefore can be predicted by time series methods; and (iii) model-based predictions that use solar dynamo models to predict the solar activity. Among the methods corresponding to the first group, there are several relationships between different parameters of the solar activity which are widely used for prediction purposes. As an example, we would highlight the Waldmeier Effect (anticorrelation between solar maximum amplitude and rise time), the Amplitude-Minimum Effect (correlation between solar minimum and maximum amplitude), the Gnevyshev Effect (correlation between the solar maximum amplitude for an odd cycle and the solar maximum amplitude of the preceding even cycle), *inter alia* (Solanki *et al*., 2002; Du and Du, 2006; Kane, 2008; Carrasco *et al*., 2016).

In this article, we revise the previous work by Wilson (1995), Wilson and Hathaway (2005), and Wilson and Hathaway (2006) which employed the spotless days as a predictor of solar activity. However, we shall use the recent version of the sunspot number (Version 2) published by SILSO (Sunspot Index and Long-Term Solar Observations, http://www.sidc.be/silso/) (Clette et al., 2016). Although we recognize the data sample size limitation (we have considered data from SC 10), we want to highlight that these are the best data available and that previous considerations of timing relationships using the previous version of this dataset have been useful in solar and solar-terrestrial studies. Wilson (1995) used the first spotless day that occurred during the decline phase of the solar cycle to predict the onset of the new cycle. Later, Wilson and Hathaway (2005) made a detailed analysis of the spotless days and their relationship with the duration and magnitude of the solar cycle. From the study of several relationships between solar cycle parameters (as, for example, the time between first spotless day after a solar maximum to the next solar minimum or the time between last spotless day occurrence after a solar minimum and the next solar maximum), Wilson and Hathaway (2005) found a systematic behaviour that clearly differed between the more recent SC 16-23 and the earlier SC 10-15. Wilson and Hathaway (2005) also noted that, around the solar minimum, the number of spotless days increases rapidly to reach a peak, and then decreases rapidly. Later, Wilson and Hathaway (2006) examined more widely this particular aspect of their previous work.

A number of reasons led us to carry out this new study about the spotless days. First, we employ Version 2 of the sunspot number instead the Version 1 used in the previous works by Wilson (1995), Wilson and Hathaway (2005), and Wilson and Hathaway (2006). Furthermore, we employ the 24-month filter defined by Hathaway (2015) to smooth the sunspot number series. This filter is preferable to the traditional 13-month running mean because the 13-month running mean does not remove the high-frequency solar activity variations (Hathaway, 2015). Double-peak maxima in the sunspot number caused by solar activity variations on timescales of 1-3 years are filtered out by the 24-month filter. Another motivation for this study is to reanalyze those relationships proposed in the previous works of Wilson (1995), Wilson and Hathaway (2005), and Wilson and Hathaway (2006) including the addition of new data, particularly the extended minimum of solar cycle 23 and the weak minimum of cycle 24. In Section 2, we present the data used and the parameters included in this study. Section 3 is devoted to presenting and discussing the results, and finally, in Section 4, we present the main conclusion to be drawn from this work.

## 2. Data

We took Version 2 of the sunspot number index that has recently been published (www.sidc.be/silso/datafiles) to revisit the use of spotless days as a predictor of solar activity. A 24-month filter proposed by Hathaway (2015) was applied to the sunspot number series. We then calculated the following parameters:

- $t_1$: elapsed time [months] from the solar minimum to the first spotless day after the solar maximum.
- $t_2$: elapsed time [months] from solar maximum to the first spotless day that occurred after the solar maximum.
- $t_3$: elapsed time [months] from the first spotless day after the solar maximum to the following solar minimum.
- $t_4$: elapsed time [months] from the last spotless day before solar maximum to the first spotless day that occurred between solar maximum and the following solar minimum.
- $t_5$: elapsed time [months] from the first spotless day that occurred between the previous solar maximum and solar minimum to the last spotless day that occurred between solar minimum and solar maximum.
- $t_6$: elapsed time [months] from the last spotless day that occurred between solar minimum and maximum to the solar maximum.
- *RM* and *Rm*: maximum and minimum sunspot number amplitudes, respectively.
- PER: elapsed time [months] between two successive solar minima.
- $NSD_{10}$, $NSD_{15}$, and $NSD_{20}$: elapsed time [months] from the first appearance of 10, 15, 20 or more spotless days per month that occurred after the solar maximum for a given solar cycle to the same for the preceding solar cycle, respectively.

- $t_{10}$, $t_{15}$, and $t_{20}$: elapsed time [months] from the solar minimum to the first appearance of 10, 15, 20 or more spotless days per month that occurred after the solar maximum of the previous solar cycle, respectively.

- $t_{10f\text{-}l}$, $t_{15f\text{-}l}$, and $t_{20f\text{-}l}$: elapsed time [months] from the first occurrence of 10, 15, 20 or more spotless days (that occurred between solar minimum for a given solar cycle and the maximum of the previous cycle) to the last occurrence of 10, 15, 20 or more spotless days (that occurred between minimum and maximum for a given solar cycle), respectively.

- $t_{10l\text{-}M}$, $t_{15l\text{-}M}$, and $t_{20l\text{-}M}$: elapsed time [months] from the last occurrence of 10, 15, 20 or more spotless days that occurred between minimum and maximum for a given solar cycle to the solar maximum, respectively.

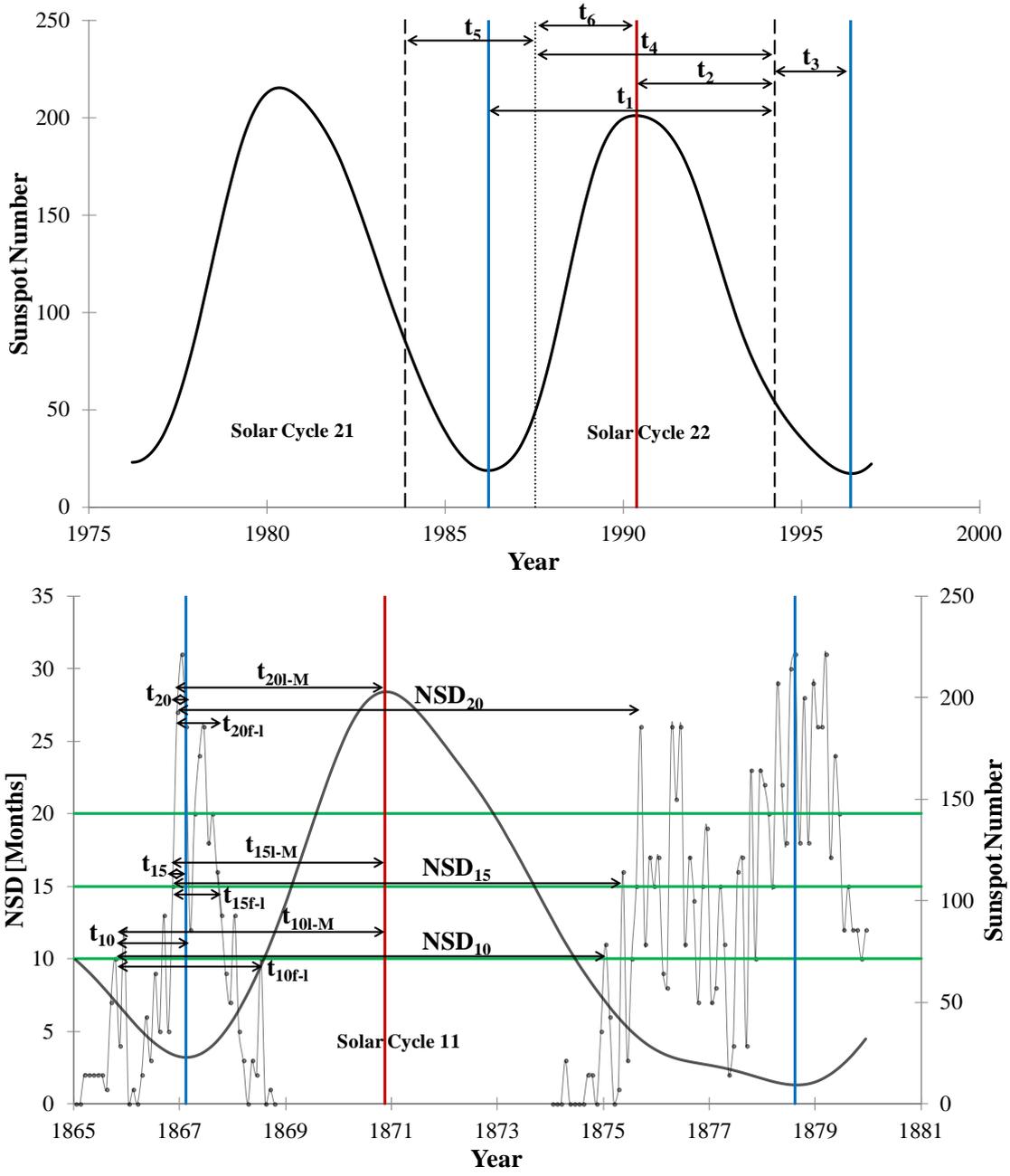

Figure 1. A schematic view with all the parameters studied in this work (see the definitions in Section 2) from the new sunspot number index (continuous line in both top and bottom panels). The *vertical solid red* and *blue lines* represent the dates for solar maxima and minima. FSD (*vertical dashed lines* in the *top panel*) is the first spotless day occurred after a solar maximum, LSD (*vertical dotted line* in the *top panel*) is the last spotless day that occurred before solar maximum, and NSD (*grey dots* in the *bottom panel*) is the elapsed time from the first appearance of 10, 15, 20 or more spotless days per month that occurred after the solar maximum to the same for the preceding solar cycle.

To clarify all of these definitions, a schematic view is given in Figure 1. In the case of duplicate sunspot number values, for example, for the solar minimum amplitude, we choose the first one in time to set the occurrence of the solar minimum. Tables 1 and 2 contain the values obtained for all the parameters described above with the new version of the sunspot number. Note that we use data just for SC 10-24 because from SC 10 there are no gaps in the series (Vaquero, 2007). Note also that some parameters cannot yet be calculated for SC 24. As mentioned above, parameters analyzed in this paper from new version of sunspot number were previously studied by Wilson (1995) ($t_1$, $t_2$, and $t_3$), Wilson and Hathaway (2005) ($t_4$, $t_5$, and $t_6$), and Wilson and Hathaway (2006) ($NSD_{10}$, $NSD_{15}$, $NSD_{20}$, $t_{10}$, $t_{15}$, $t_{20}$, $t_{10f\text{-}l}$, $t_{15f\text{-}l}$, $t_{20f\text{-}l}$, $t_{10l\text{-}M}$, $t_{15l\text{-}M}$, and $t_{20l\text{-}M}$) using the old sunspot number (version 1). However, the main cause of the differences in the values of parameters involving the timing of solar maximum found in this work with respect to those of the previous work is the 24-month Gaussian filter proposed by Hathaway (2015) used in this study to smooth the sunspot number series instead of the traditional 13-month smoothing used in those earlier works.

Table 1. Values of the parameters *RM*, *Rm*, PER, $t_1$, $t_2$, $t_3$, $t_4$, $t_5$, and $t_6$ for SC 10-24 from the new sunspot number series.

| Solar Cycle | RM | Rm | PER | $t_1$ | $t_2$ | $t_3$ | $t_4$ | $t_5$ | $t_6$ |
|---|---|---|---|---|---|---|---|---|---|
| 10 | 174.9 | 13.2 | 132 | 68 | 19 | 64 | 42 | 107 | 23 |
| 11 | 203.0 | 22.9 | 138 | 75 | 30 | 63 | 46 | 93 | 16 |
| 12 | 107.9 | 9.3 | 130 | 77 | 14 | 53 | 16 | 124 | 2 |
| 13 | 135.7 | 10.5 | 147 | 77 | 26 | 70 | 47 | 83 | 21 |
| 14 | 99.4 | 7.6 | 135 | 61 | 5 | 74 | 15 | 116 | 10 |
| 15 | 147.7 | 5.0 | 125 | 88 | 28 | 37 | 42 | 120 | 14 |
| 16 | 119.4 | 15.3 | 123 | 88 | 33 | 35 | 50 | 75 | 17 |
| 17 | 180.4 | 12.5 | 125 | 99 | 47 | 26 | 76 | 58 | 29 |
| 18 | 202.3 | 23.0 | 121 | 83 | 35 | 38 | 63 | 46 | 28 |
| 19 | 266.2 | 16.2 | 128 | 93 | 45 | 35 | 73 | 58 | 28 |
| 20 | 150.9 | 21.9 | 137 | 105 | 52 | 32 | 83 | 57 | 31 |
| 21 | 215.2 | 22.9 | 120 | 92 | 42 | 28 | 76 | 48 | 34 |
| 22 | 200.9 | 18.6 | 122 | 97 | 49 | 25 | 81 | 44 | 32 |
| 23 | 168.8 | 17.1 | 149 | 92 | 34 | 57 | 72 | 45 | 38 |
| 24 | 100.8 | 5.4 | - | 69 | 5 | - | 35 | 91 | 30 |

Table 2. Values of the parameters $NSD_{10}$, $NSD_{15}$, $NSD_{20}$, $t_{10}$, $t_{15}$, $t_{20}$, $t_{10f-l}$, $t_{15f-l}$, $t_{20f-l}$, $t_{10l-M}$, $t_{15l-M}$, and $t_{20l-M}$ for SC 10-24 from the new sunspot number series.

| Solar Cycle | $NSD_{10}$ | $NSD_{15}$ | $NSD_{20}$ | $t_{10}$ | $t_{15}$ | $t_{20}$ | $t_{10f-l}$ | $t_{15f-l}$ | $t_{20f-l}$ | $t_{10l-M}$ | $t_{15l-M}$ | $t_{20l-M}$ |
|---|---|---|---|---|---|---|---|---|---|---|---|---|
| 10 | 132 | 145 | 140 | 16 | 16 | 10 | 30 | 29 | 18 | 35 | 36 | 41 |
| 11 | 111 | 102 | 105 | 16 | 3 | 2 | 33 | 10 | 8 | 28 | 38 | 39 |
| 12 | 142 | 138 | 134 | 43 | 39 | 35 | 59 | 51 | 45 | 47 | 51 | 53 |
| 13 | 140 | 153 | 153 | 31 | 31 | 31 | 50 | 43 | 43 | 32 | 39 | 39 |
| 14 | 145 | 135 | 142 | 38 | 25 | 25 | 62 | 40 | 40 | 32 | 41 | 41 |
| 15 | 136 | 141 | 139 | 28 | 25 | 18 | 50 | 45 | 33 | 38 | 40 | 45 |
| 16 | 114 | 121 | 123 | 17 | 9 | 4 | 37 | 29 | 14 | 35 | 35 | 45 |
| 17 | 147 | 136 | 129 | 26 | 11 | 4 | 42 | 24 | 9 | 36 | 39 | 47 |
| 18 | 113 | 109 | 118 | 4 | 0 | 0 | 10 | 4 | 4 | 42 | 44 | 44 |
| 19 | 136 | 137 | 128 | 12 | 12 | 3 | 26 | 25 | 10 | 34 | 35 | 41 |
| 20 | 130 | 129 | 144 | 4 | 3 | 3 | 13 | 0 | 0 | 44 | 56 | 56 |
| 21 | 117 | 125 | 114 | 11 | 11 | -4 | 19 | 19 | 0 | 42 | 42 | 46 |
| 22 | 123 | 125 | 128 | 14 | 6 | 2 | 25 | 17 | 5 | 39 | 39 | 47 |
| 23 | 130 | 133 | 127 | 13 | 3 | -4 | 27 | 8 | 1 | 46 | 55 | 55 |
| 24 | - | - | - | 32 | 19 | 18 | 51 | 32 | 30 | 45 | 51 | 52 |

## 3. Analysis

### 3.1. Predicting the timing of solar minimum

Wilson (1995) proposed the "first spotless day" (defined as the first sunspot record equal to zero during the decline of the solar cycle) as a predictor for the occurrence of the minimum of the following solar cycle. In that work, Wilson predicted the solar minimum of SC 22 based on the linear and modal secular fits of the parameter $t_3$, and from correlations with respect to observed values of $t_1$ and $t_2$. Unlike the present work, Wilson used the old version of the international sunspot number and the traditional 13-month running mean to make the predictions, and also considered the SC 9.

Figure 2 shows the temporal evolution of the parameter $t_3$. Wilson (1995) showed a decrease with time for $t_3$, while in this work that decrease is broken with SC 23 since the parameter $t_3$ for this cycle is significantly greater than for the eight earlier solar cycles. Regardless of the methodological approach, that is the main reason why the correlation obtained in the present work is weaker than those found by Wilson (1995). Using the linear fit of Figure 2, this method does not predict well the parameter $t_3$ for SC 23 since the estimated value would be equal to 28 months while the observed value is equal to 57 months, i.e., twice the estimated value. If we perform the same calculation discarding SC 23, the estimated $t_3$ for SC 23 would be even lower (18 months), but the correlation stronger ($r = -0.82$; $p$-value $< 0.001$) than that obtained considering all the data. Therefore, although the value of the correlation coefficient is high ($r = -0.64$; $p$-value $= 0.014$), it would be unsuitable to make a prediction for the occurrence of the following solar minimum based on this linear fit of the parameter $t_3$. Wilson (1995) provided another interpretation of the variation of the parameter $t_3$ through a "modal effect". In our work, it can also be seen in Figure 2 that SC 10-14 and 23 seem to show a different behaviour to that of SC 15-22. Defining long and short solar cycles if the period for each cycle is greater or less than the mean value of all the solar cycle periods (130.9 months), respectively, we note that: (i) the first subset (SC 10-14 and 23) is composed of long solar cycles, except for SC 12 although its period (130 months) is very close to the mean value of all periods; and (ii) the second one (SC 15-22) consists of short solar cycles, except for SC 20 which is a long cycle. We obtained the mean value and standard deviation of $t_3$ for each subset: (i) 32.0 ± 5.1 months for SC 15-22, and (ii) 63.5 ± 7,8 months for SC 10-14 and 23. Assuming that, on the one hand, the behaviour of SC 24 will be similar to SC 10-14 and 23, and since the first spotless day for the decline phase corresponding to SC 24 occurred in July 2014, the minimum for SC 25 would be expected around 2019 and the first half of 2020.

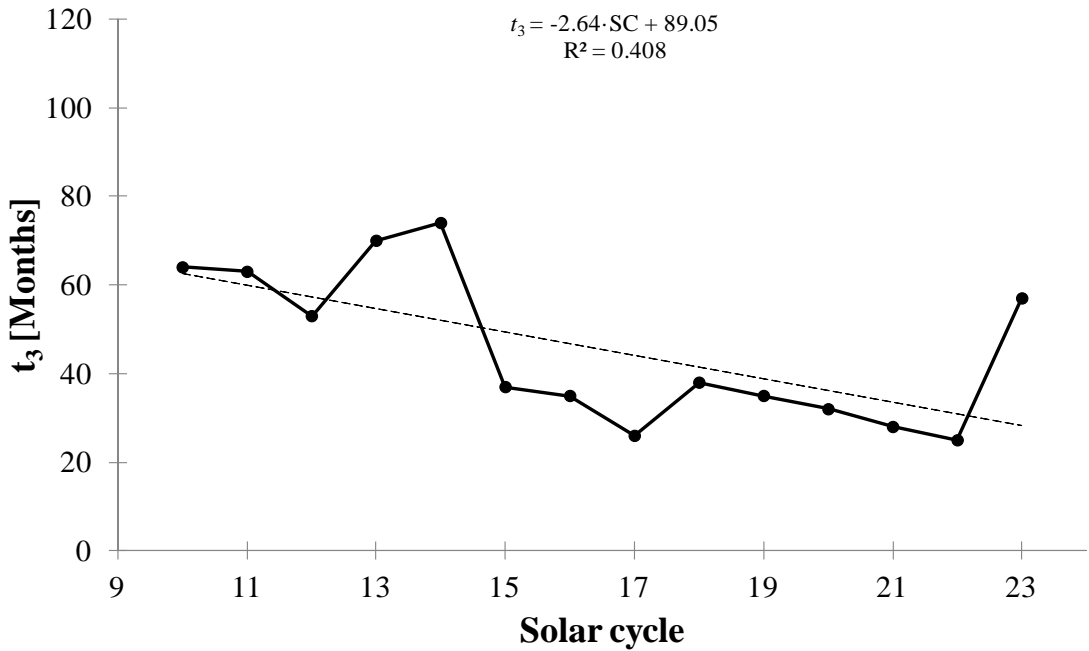

Figure 2. Temporal evolution of the parameter $t_3$ as a function of the solar cycle number for SC 10-23. The *dashed line* represents the best linear fit. Its equation and *R*-squared coefficient are shown.

Wilson (1995) also proposed estimates of the occurrence of the solar minimum from the correlation of the parameter $t_3$ with $t_2$ and $t_1$, independently. Figure 3 depicts the best linear fits between these parameters. Both fits have a good level of correlation and high significance: (i) $r = -0.85$, *p*-value < 0.001 for $t_3$ versus $t_1$; and (ii) $r = -0.82$, *p*-value <

0.001 for $t_3$ versus $t_2$. This implies that cycles with large $t_1$ (elapsed time between the solar minimum and the first spotless day during the decline phase) and $t_2$ (elapsed time between the solar maximum and the first spotless day during the decline phase) have shorter $t_3$ (elapsed time between the first spotless day during the decline phase and the following solar minimum) and vice versa. Although it seems clear that the data has this pattern, it is difficult to make a good estimate for the occurrence of a particular minimum from these linear fits. For example, according to this analysis, the estimated values for SC 23 from the $t_3$ versus $t_1$ and $t_2$ linear fits are 38 and 44 months, respectively. These values are far from the observed value for $t_3$ which is equal to 57 months. This means differences between the observed and estimated values greater than one year: 13 and 19 months for $t_2$ and $t_1$, respectively. Removing SC 23 from the analysis improves the correlation coefficients and the significance level ($r = -0.91$, $p$-value $< 0.001$ for the $t_1$ linear fit, and $r = -0.84$, $p$-value $< 0.001$ for the $t_2$ linear fit). However, the estimated values are even lower (36 months from the $t_1$ linear fit, and 43 months from the $t_2$ linear fit) than those with SC23 considered in the analysis. Based on these fits, the $t_3$ estimated values for SC 24 from the $t_1$ and $t_2$ linear fits are 65 and 74 months, respectively. These results imply that the SC 25 minimum will occur around the final part of the year 2019 or during 2020.

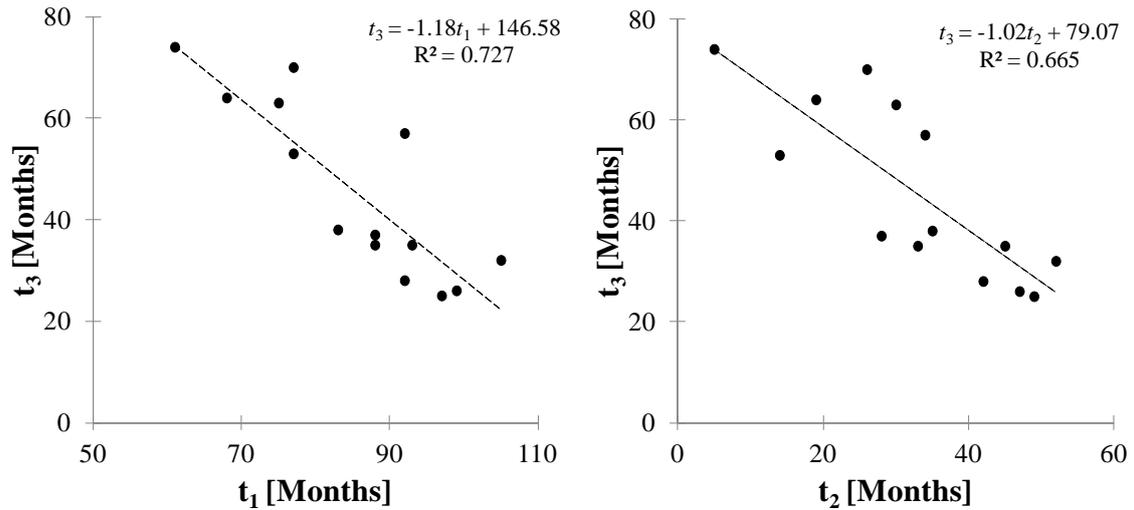

Figure 3. Linear fits between the parameters $t_3$ and $t_1$ (left panel) and $t_3$ and $t_2$ (right panel) for SC 10-23. The *dashed lines* represent the best linear fits. Their equations and *R*-squared coefficients are shown.

### 3.2. Time series and correlations for parameter $t_4$.

Wilson and Hathaway (2005) used the spotless days in relation to the timing and size of the solar cycle to the study of the characteristics of sunspot cycles and they found that the behavior of the most recent cycles, SC 16–23, was markedly different from the earlier cycles, SC 10–15. Moreover, Wilson and Hathaway (2005) pointed out that, considering SC 10-23, the parameter $t_4$ shows a substantial systematic increase. Figure 4 represents the temporal evolution of this parameter for SC 10-24. It shows that, including SC 24, systematic increase no longer exists since the *$t_4$* value for SC 24 is equal to 35 months which it is the third lowest value of the whole series. Analogously to the parameter *$t_3$* in Figure 2, this *$t_4$* sharp drop in Figure 4 for SC 24 might represent a mode change that may persist for several solar cycles.

Figure 5 depicts scatterplots of $t_4$ against $t_3$, $t_5$, and $t_6$. It can be seen that the three linear fits have high values for both the correlation coefficients and the significance level, namely: (i) $r = -0.71$, $p$-value $= 0.003$ for $t_4$ versus $t_3$, (ii) $r = -0.91$, $p$-value $< 0.001$ for $t_4$ versus $t_5$, and (iii) $r = -0.84$, $p$-value $< 0.001$ for $t_4$ versus $t_6$. In general, when the values of $t_3$ and $t_5$ are high then the $t_4$ values are low, and vice versa. In contrast, the parameters $t_4$ and $t_6$ have the same behavior: when the $t_6$ values are high (low) the $t_4$ values are high (low). Despite the strong correlations, estimation of the parameter $t_4$ from these linear fits can fail if one particularizes for a single solar cycle. It can be seen that there are several points far from the linear fits, especially for the fits with $t_3$ and $t_6$. We would highlight the strong correlation between the parameters $t_4$ and $t_5$. However, it would be difficult to defend the reliability of any prediction based on these correlations. For example, for the current SC 24 and taking into account those linear fits, the estimated value for $t_4$ is equal to 45 months while the observed value is 35 months. This implies an absolute error approximately equal to 20 %. In other cases, such as SC 20 for example, the error in the estimates (69 months) with respect to the observed value (83 months) is greater than one year. Therefore, this linear fit should be used with caution.

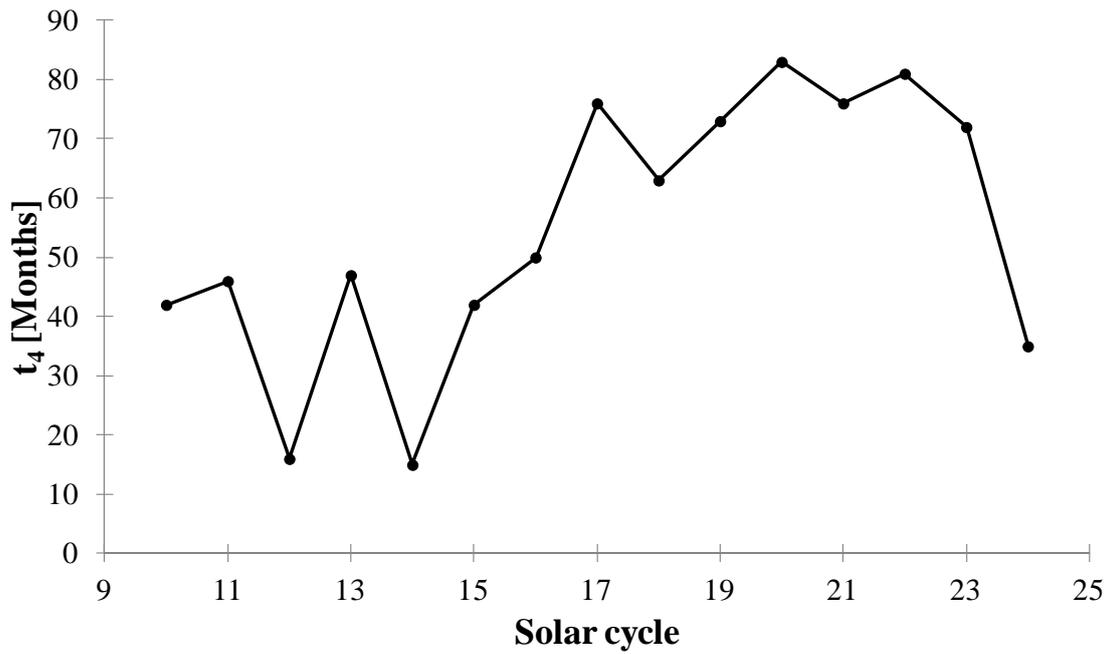

Figure 4. Temporal evolution of the parameter $t_4$ as a function of the solar cycle number for SC 10-24.

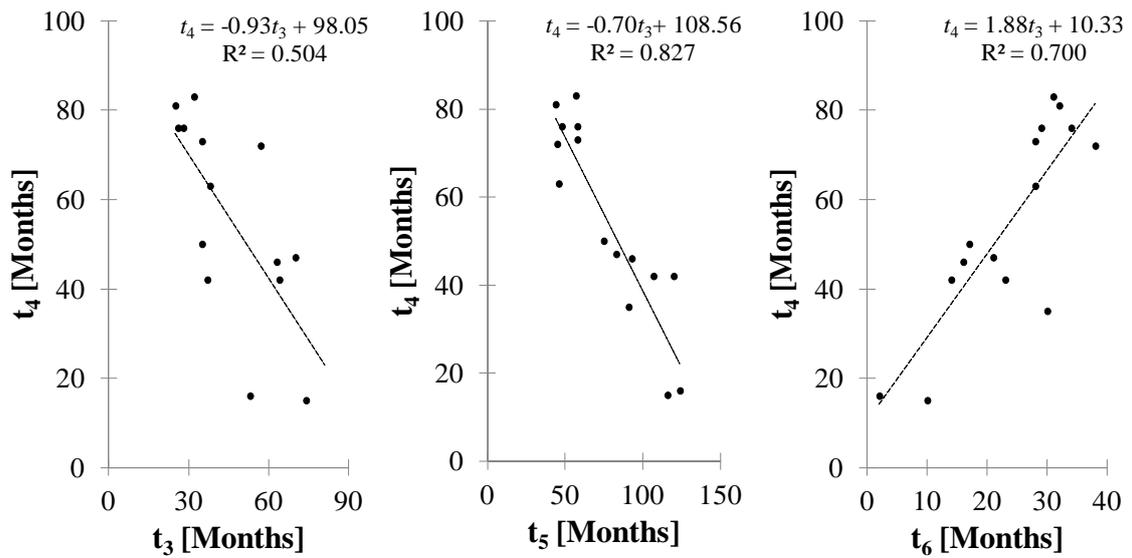

Figure 5. Linear fits between the parameters $t_4$ and $t_3$ (left panel), $t_4$ and $t_5$ (*middle panel*), and $t_4$ and $t_6$ (*right panel*) for SC 10-24. The *dashed lines* represent the best linear fits. Their equations and *R*-squared coefficients are shown.

### 3.3. $NSD_{20}$ as a guage of solar cycle length

Wilson and Hathaway (2006) proposed the parameter $NSD_{20}$ as a reliable gauge of the time of onset of the following solar cycle, arguing that when the $NSD_{20}$ value for a given solar cycle is equal to or lower than the $NSD_{20}$ median value considering all the solar cycles then the PER value of that cycle is equal to or shorter than the PER median value taking into account all the solar cycles (the only exception being for SC 11), and when the $NSD_{20}$ value is greater than its median value, the PER value is greater than the PER median except for the case of the SC 15. Figure 6 represents the variation of $NSD_{10}$, $NSD_{15}$, $NSD_{20}$, and PER. It shows that this statement is not valid for SC 23 which could not be included in the analysis of Wilson and Hathaway (2006). According to that statement, the PER value of SC 23 should be equal to or less than 129 months (the PER median value) since $NSD_{20}$ for SC 23 is less than the median value. However, the PER value of SC 23 (149 months) is well above the median value considering all the cycles (129 months). In our case, the $NSD_{20}$ value for SC 17 (129 months) is slightly greater than the $NSD_{20}$ median value (128.5 months), but the SC 17 PER value (125 months) is less than the PER median value (129 months). Moreover, it can be seen that SC 23 has broken the trend of SC 15-22 (except for SC 20) for which the value of the period was below the median value (129 months). SC 23 is the longest cycle (149 months) in Table 1.

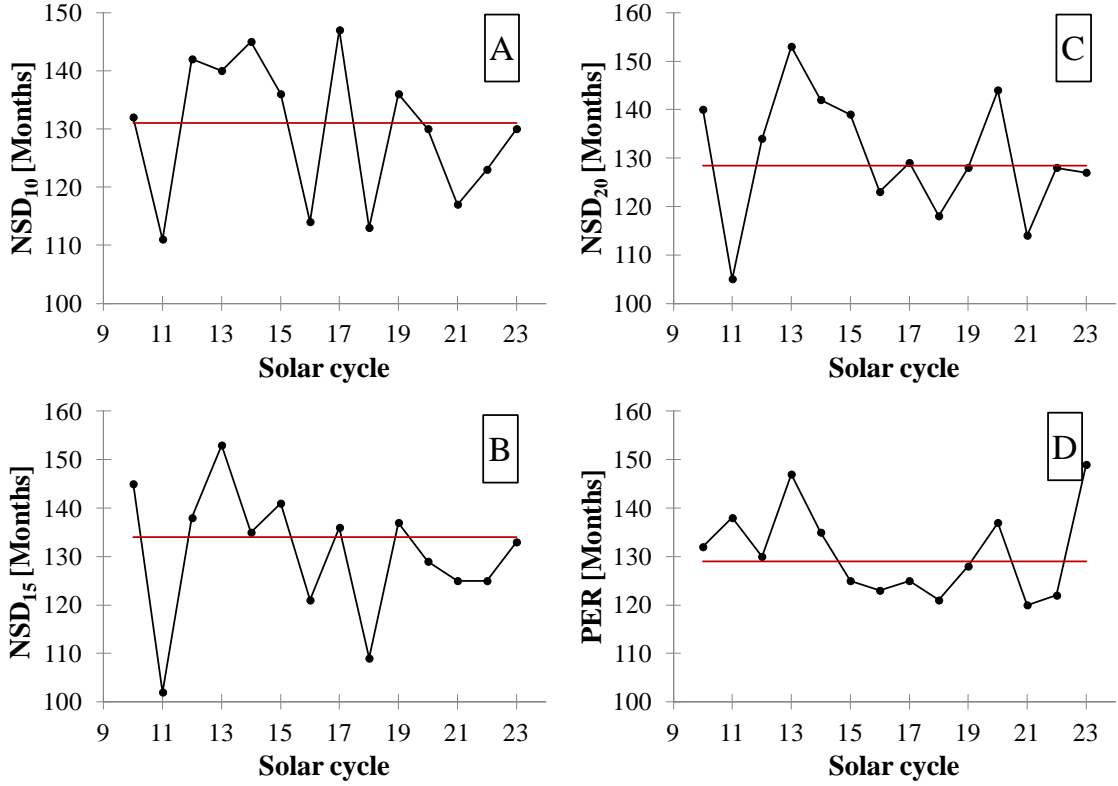

Figure 6. Temporal evolution of the parameters $NSD_{10}$ (*panel A*), $NSD_{15}$ (*panel B*), $NSD_{20}$ (*panel C*), and PER (*panel D*) as functions of the solar cycle number for SC 10-23. The *horizontal line* in each panel represents the median value of the data, and is equal to 131, 134, 128.5, and 129 months, respectively.

Figure 7 displays PER versus $NSD_{10}$ (left panel), $NSD_{15}$ (middle panel), and $NSD_{20}$ (right panel). Wilson and Hathaway (2006), on the basis of the traditional smoothed sunspot number, found that the PER values were clearly distributed into two non-overlapping groups, one comprising solar cycles with short periods and the other solar cycles with longer periods. Unlike the work of Wilson and Hathaway (2006), we do not find this distribution on the basis of the 24-month filter. Moreover, Wilson and Hathaway (2006) indicated that if $NSD_{20}$ for SC 23 were equal to or lower than its median value then SC23 would be more likely to have a short period. However, according to our results, SC 23 is a long cycle despite its $NSD_{20}$ value being lower than the median, and therefore does not fulfill the result expected by Wilson and Hathaway (2006). If we discard the two outlier points (SC 11 and 23) in the scatter plot for PER versus $NSD_{20}$, the remaining points can be fitted with polynomials giving high values of both the correlation coefficient and the significance level: $PER = 0.02(NSD_{20})^2 - 3.77\ NSD_{20} + 335$, $r = 0.95$, $p$-valor $< 0.001$. Therefore, in general, it seems that low (high) values of the solar cycle PER are related to low (high) values of $NSD_{20}$.

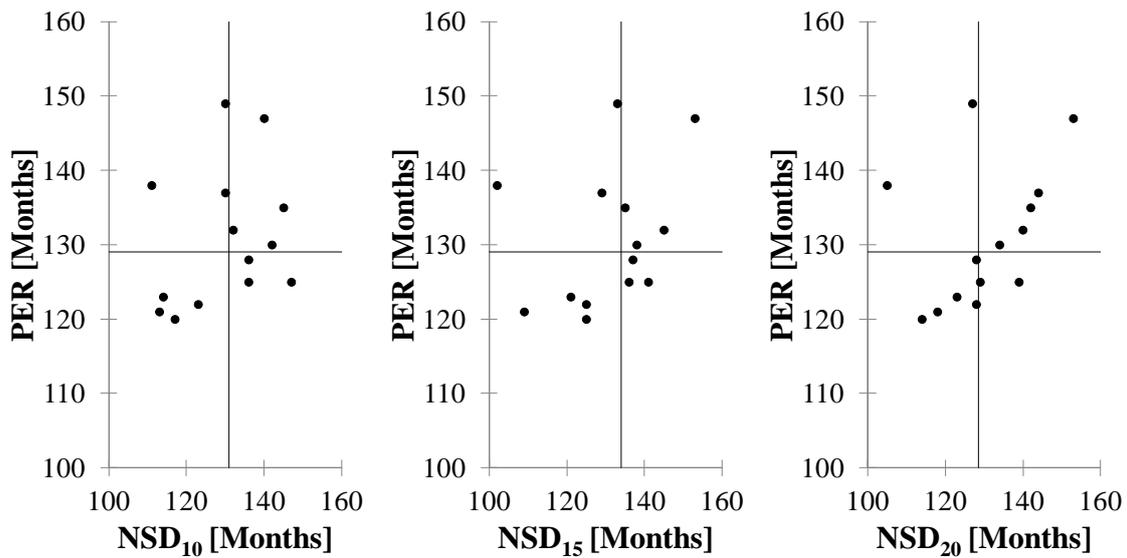

Figure 7. Scatter plot for PER versus $NSD_{10}$ (*left panel*), $NSD_{15}$ (*middle panel*), and $NSD_{20}$ (*right panel*). The *horizontal* and *vertical lines* represent the median values for PER and NSD, respectively.

## 3.4. Time series and correlations for $t_{10}$, $t_{15}$, and $t_{20}$.

Wilson and Hathaway (2006) predicted that, once $NSD_{10f}$, $NSD_{15f}$, and $NSD_{20f}$ occur, the minimum of SC 24 would probably not surpass in time the median value of these parameters. Figure 8 represents the cyclical variation for $t_{20}$ (left panel), $t_{15}$ (middle panel), and $t_{10}$ (right panel). These parameters measure the elapsed time between $NSD_{10f}$, $NSD_{15f}$, and $NSD_{20f}$ and the solar minimum. It can be seen that SC 24 does not have the same behaviour as the previous SC 16-23, and one appreciates a significant increase of the values of $t_{20}$, $t_{15}$, and $t_{10}$ in that solar cycle. In our results, it can be seen that, in general, the values corresponding to $t_{20}$, $t_{15}$, and $t_{10}$ for SC 10-15 and 24 (except for SC 11) are significantly greater than those for SC 16-23 (except for SC 16 and 17 for $t_{10}$).

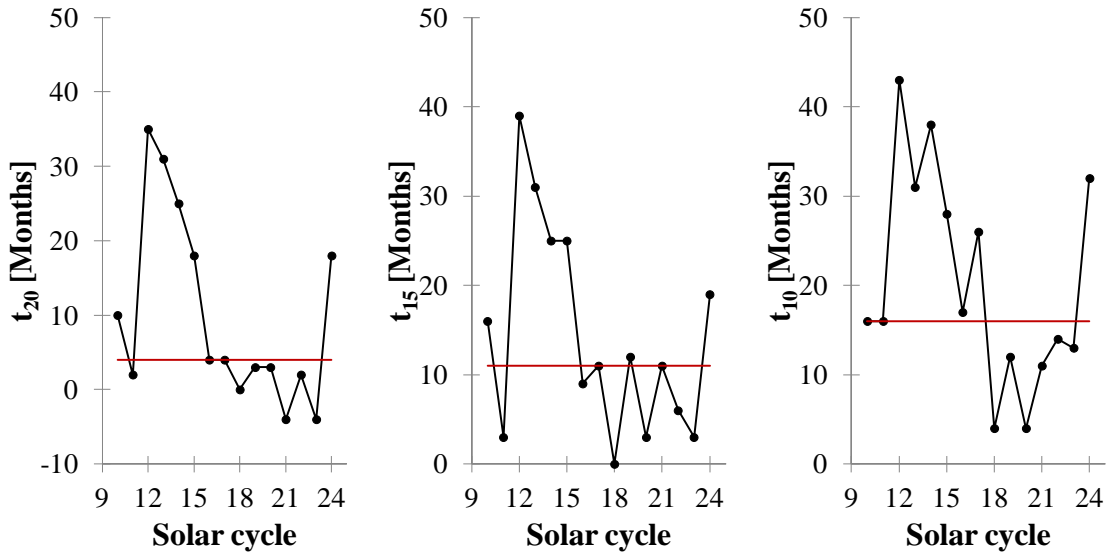

Figure 8. Temporal evolution of the parameters $t_{20}$ (*left panel*), $t_{15}$ (*middle panel*), and $t_{10}$ (*right panel*) as functions of the solar cycle number for SC 10-24. The *horizontal line* in each panel represents the median value of the data, equal to 4, 11, and 16 months, respectively.

Figure 9 displays the parameters *RM* and *Rm* against $t_{10}$, $t_{15}$, and $t_{20}$. We obtain the same behavior for these data as Wilson and Hathaway (2006), i.e., these linear fits suggest that large values of $t_{10}$, $t_{15}$, and $t_{20}$ seem to be associated with small values of *RM* and *Rm* and *vice versa*. It can be seen that for the linear fit between *Rm* and $t_{20}$ all data with $t_{20}$ values greater than the median (4 months) have *Rm* values less than its median (15.3), and all data with $t_{20}$ values greater than 4 months (the $t_{20}$ median) have *Rm* values greater than 15.3 (the *Rm* median). The most significant correlation found is for the linear fits between *Rm* and $t_{10}$ ($r = -0.85$; $p$-value $< 0.001$). Again, although the correlation is strong, the estimated values are sometimes far from the observed values. For example, for the current SC 24 the estimated value for *Rm* according to the linear fit is equal to 9.5 and the observed value is equal to 5.4, which it is approximately half the estimated value. We want to emphasize that the correlations obtained in this work with the new sunspot number, a 24-month filter, and for SC 10-24 are significantly stronger than the correlation obtained by Wilson and Hathaway (2006) with the old sunspot number, the traditional smoothed sunspot number, and for SC 10-23.

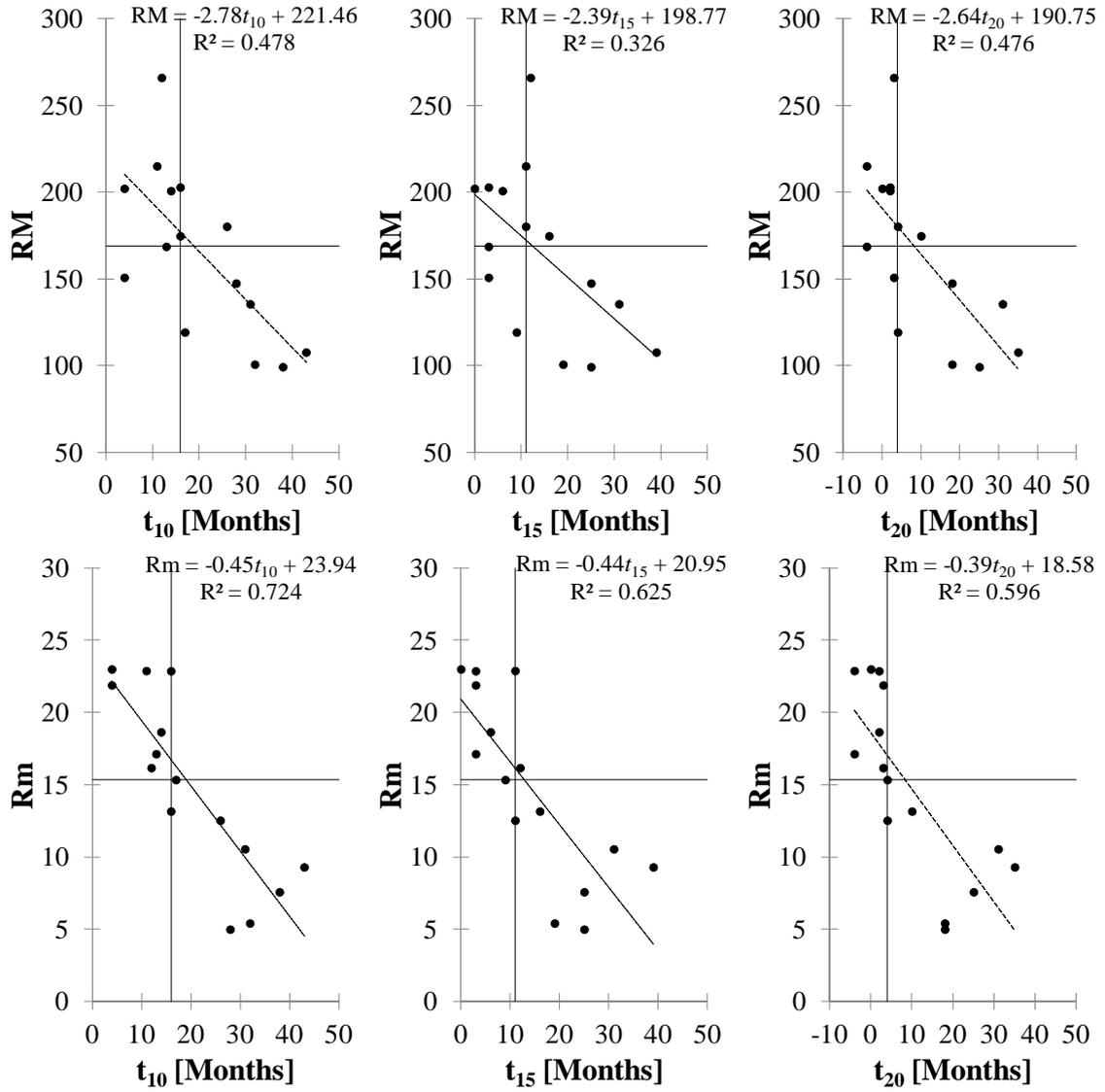

Figure 9. Linear fits of the parameters *RM* (*upper panels*) and *Rm* (*bottom panels*) versus $t_{10}$ (*left panels*), $t_{15}$ (*middle panels*), and $t_{20}$ (*right panels*) for SC 10-24. The *dashed lines* represent the best linear fits. Their equations and *R*-squared coefficients are shown. The *horizontal* and *vertical lines* represent the median values for *RM* or *Rm* and $t_{10}$, $t_{15}$, or $t_{20}$, respectively.

**3.5. Time series and correlations for $t_{10f\text{-}l}$, $t_{15f\text{-}l}$, $t_{20f\text{-}l}$, $t_{10l\text{-}M}$, $t_{15l\text{-}M}$, and $t_{20l\text{-}M}$.**

Figure 10 shows the temporal evolution of the parameters $t_{20f\text{-}l}$, $t_{15f\text{-}l}$, and $t_{10f\text{-}l}$. It can be seen that, omitting the current SC 24, the values for these parameters of the last solar cycles are below the median value. Wilson and Hathaway (2006) pointed out that if that trend continued then the values for $t_{10f\text{-}l}$, $t_{15f\text{-}l}$, and $t_{20f\text{-}l}$ corresponding to SC 24 should be not greater than their median values. However, the values of $t_{10f\text{-}l}$, $t_{15f\text{-}l}$, and $t_{20f\text{-}l}$ for SC 24 increase significantly. Therefore, as was the case with the parameters $t_{10}$, $t_{15}$, and $t_{20}$, the behavior of SC 24 differs from that of the previous six or seven solar cycles. Figure 11 depicts the temporal variation of $t_{20l\text{-}M}$, $t_{15l\text{-}M}$, and $t_{10l\text{-}M}$ for SC 10-24. Wilson and Hathaway (2006) indicated that there are hints of a four-cycle variation in these parameters. Although our results for $t_{20l\text{-}M}$, $t_{15l\text{-}M}$, and $t_{10l\text{-}M}$ are not similar to those obtained by Wilson and Hathaway (2006), there are also hints of a small cyclical variation approximately equal to four cycles for $t_{15l\text{-}M}$ and $t_{10l\text{-}M}$, but not at all clearly for $t_{20l\text{-}M}$.

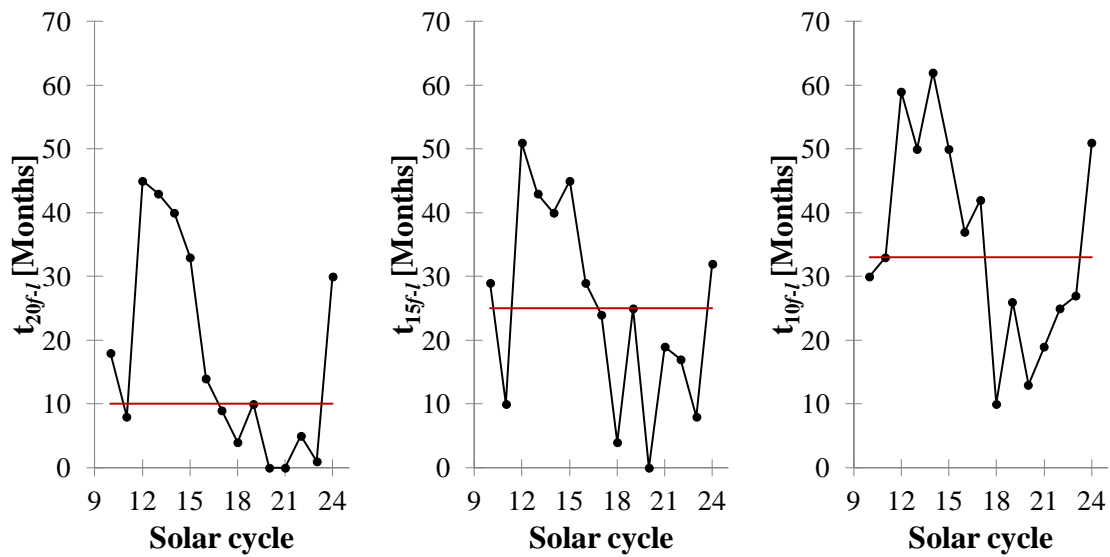

Figure 10. Temporal evolution of the parameters $t_{20f\text{-}l}$ (*left panel*), $t_{15f\text{-}l}$ (*middle panel*), $t_{10f\text{-}l}$ (*right panel*) as functions of the solar cycle number for SC 10-24. The *horizontal*

*line* in each panel represents the median value of the data, equal to 10, 25, and 33 months, respectively.

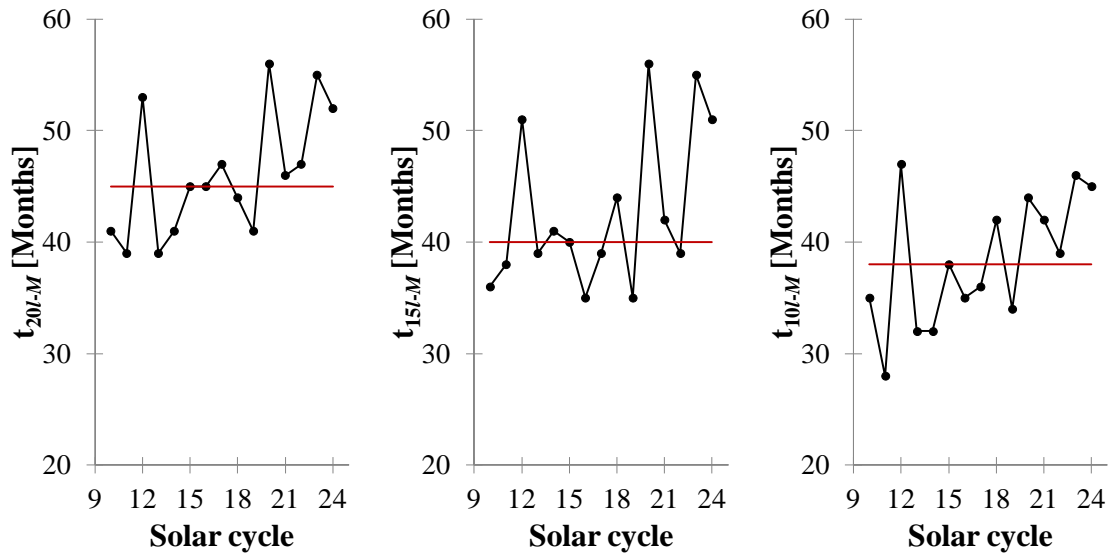

Figure 11. Temporal evolution of the parameters $t_{20l\text{-}M}$ (*left panel*), $t_{15l\text{-}M}$ (*middle panel*), and $t_{10l\text{-}M}$ (*right panel*) as functions of the solar cycle number for SC 10-24. The *horizontal line* in each panel represents the median value of the data, equal to 45, 40, and 38 months, respectively.

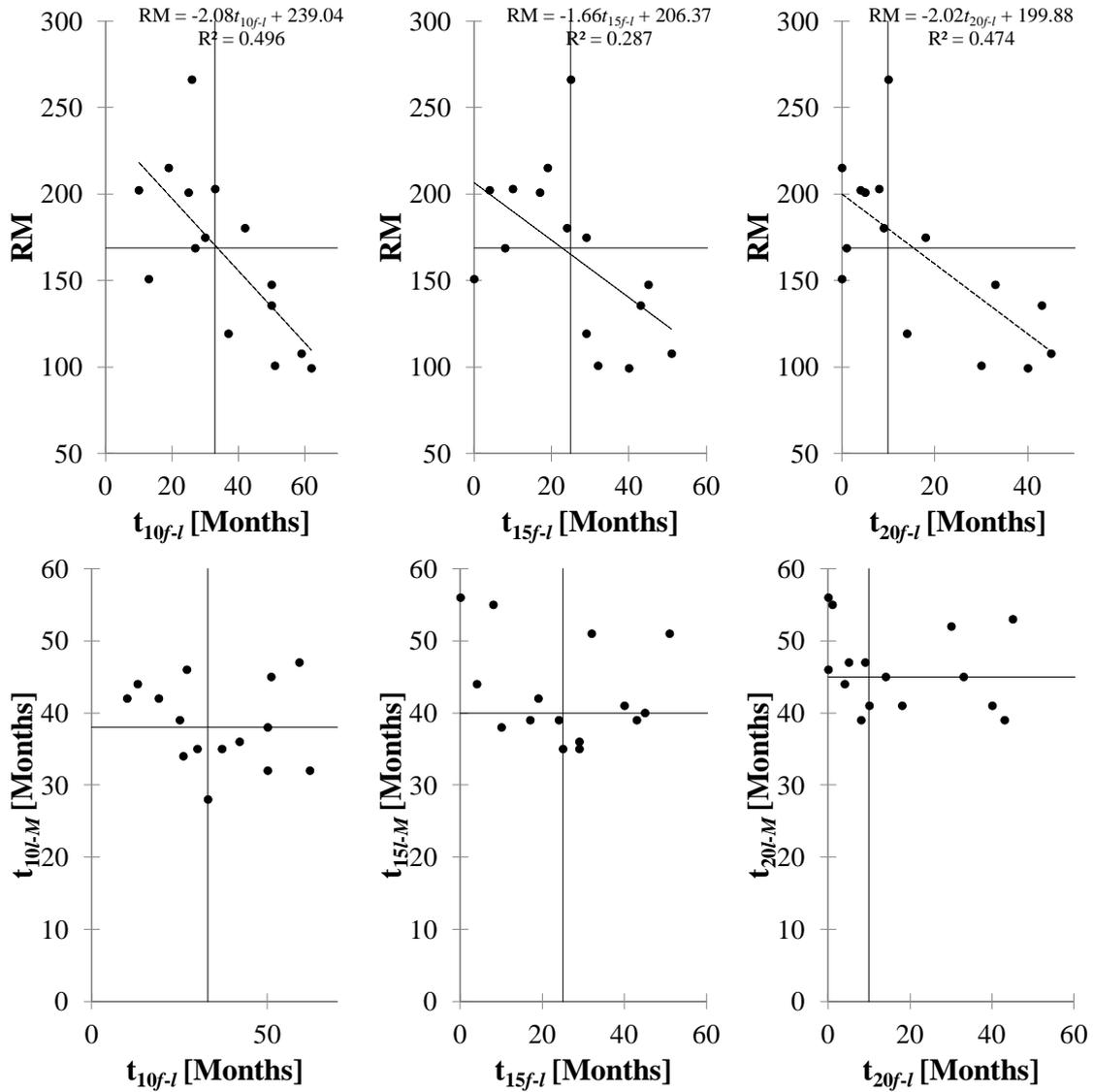

Figure 12. Linear fits between the parameters *RM* (*upper panels*) and $t_{10l\text{-}M}$, $t_{15l\text{-}M}$, and $t_{20l\text{-}M}$ (*bottom panels*) versus $t_{10f\text{-}l}$ (left panels), $t_{15f\text{-}l}$ (middle panels), and $t_{20f\text{-}l}$ (*right panels*) for SC 10-24. The dashed lines in the upper panels represent the best linear fits. Their equations and *R*-squared coefficients are shown. The *horizontal* and *vertical lines* represent the median values for *RM* and $t_{10l\text{-}M}$, $t_{15l\text{-}M}$, or $t_{20l\text{-}M}$, and $t_{10f\text{-}l}$, $t_{15f\text{-}l}$, or $t_{20f\text{-}l}$, respectively.

Figure 12 represents *RM* (upper panels) and $t_{10l\text{-}M}$, $t_{15l\text{-}M}$, and $t_{20l\text{-}M}$ (bottom panels) against $t_{10f\text{-}l}$, $t_{15f\text{-}l}$, and $t_{20f\text{-}l}$. We also obtain the same behaviour as Wilson and Hathaway (2006): data more normally distributed for the fits between $t_{10l\text{-}M}$, $t_{15l\text{-}M}$, and $t_{20l\text{-}M}$ and $t_{10f\text{-}l}$, $t_{15f\text{-}l}$, and $t_{20f\text{-}l}$, and a significant correlation for *RM versus* $t_{10f\text{-}l}$, $t_{15f\text{-}l}$, and $t_{20f\text{-}l}$. Therefore, small $t_{10f\text{-}l}$, $t_{15f\text{-}l}$, and $t_{20f\text{-}l}$ values seem to be associated with large *RM* values. We note that we obtain correlation coefficients values greater than those found by Wilson and Hathaway (2006) for the *RM* versus $t_{10f\text{-}l}$ ($r = -0.70$; *p*-value = 0.003) and $t_{20f\text{-}l}$ ($r = -0.69$; *p*-value = 0.005) linear fits, but smaller versus $t_{15f\text{-}l}$ ($r = -0.54$; *p*-value = 0.040). Again, although we found strong correlations, estimates for future solar cycles may not be accurate. As example, the estimated *RM* value according to the two best linear fits for the current SC 24 is 132.8 (for the $t_{10f\text{-}l}$ linear fit) and 139.3 (for the $t_{20f\text{-}l}$ linear fit), while the observed value for this parameter is equal to 100.8. Thus, the relative error for these values is approximately equal to 30 %, and we would also note that there are data with greater relative errors (for example, SC 16 with 35-40 %).

**4. Summary and Conclusions**

We have revised previous work (Wilson, 1995; Wilson and Hathaway, 2005; Wilson and Hathaway, 2006) in which spotless days were used as predictors of solar activity. Unlike those studies, we used the new recently published version of the sunspot number (www.sidc.be/silso/datafiles) together with a 24-month filter in order to smooth this series. The following parameters (defined in Section 2) were studied: *t*1, *t*2, *t*3, *t*4, *t*5, *t*6, *RM*, *Rm*, PER, $NSD_{10}$, $NSD_{15}$, $NSD_{20}$, $t_{10}$, $t_{15}$, $t_{20}$, $t_{10f\text{-}l}$, $t_{15f\text{-}l}$, $t_{20f\text{-}l}$, $t_{10l\text{-}M}$, $t_{15l\text{-}M}$, and $t_{20l\text{-}M}$.

In general, our results were similar to those found in the previous work in terms of correlation coefficient values and behaviour of the data. However, it was found that the solar activity predictions made in those previous works for the current SC 24 or the preceding SC 23 have not been fulfilled, despite the strong correlations between some of the parameters studied. The strongest correlations were found for the linear fits between the parameters $t3$ and $t1$ ($r = -0.85$, $p$-value $< 0.001$), $t4$ and $t5$ ($r = -0.91$, $p$-value $< 0.001$), and $Rm$ and $t_{10}$ ($r = -0.85$, $p$-value $< 0.001$). Nevertheless, the observed values show that the estimated values given by these linear fits are not always good predictions since the relative error can reach up to 30-40 %.

Particularly, we have proved that the decrease and increase of the parameters $t_3$ and $t_4$ indicated by Wilson (1995) and Wilson and Hathaway (2005), respectively, have been broken with the addition of new data of sunspot number index (see Figures 2 and 4). However, it might be a modal behavior in these parameters which persist during several solar cycles such as pointed out those previous works. From the modal behavior of $t_3$ and assuming that the behaviour of SC 24 will be similar to SC 10-14 and 23, the minimum for SC 25 would be expected around 2019 and the first half of 2020. Moreover, the parameters $t_3$ and $t_4$ present a strong correlation with $t_1$ and $t_2$ in the case of $t_3$ ($r = -0.85$, $p$-value $< 0.001$ and $r = -0.82$, $p$-value $< 0.001$, respectively) (see Figure 3), and $t_3$, $t_5$ and $t_6$ ($r = -0.71$, $p$-value $= 0.003$; $r = -0.91$, $p$-value $< 0.001$; and $r = -0.84$, $p$-value $< 0.001$, respectively) in the case of $t_4$ (see Figure 5). However, despite the strong correlation, we note that some estimated values have a large relative error with respect to the observed values.

It can be seen in Figure 6 that when the $NSD_{20}$ value for a given solar cycle is greater (equal to or lower) than the $NSD_{20}$ median value then the PER value for that cycle is greater (equal to or shorter) than the PER median value (the only exceptions are SC 11 and 15). Again, SC 23 represents a mode change with respect to the seven previous solar cycles. According to this fact, the PER value of SC 23 should be equal to or less than 129 months (the PER median value). But we can see that the PER value of SC 23 is equal to 149 months, i.e., it is the longest solar cycle of all solar cycles considered in this study (from SC 10). Moreover, from the fit between PER *versus* $NSD_{10}$, $NSD_{15}$, and $NSD_{20}$, we have obtained that in general, it seems that low values of the PER are associated to low values of $NSD_{20}$ and high values of the first one with high values of the second one (see Figure 7). If we discard the two outlier points (SC 11 and 23), the fit between PER and $NSD_{20}$ give a high value of the correlation coefficient: $r = 0.95$, *p*-value $< 0.001$.

Regarding the analysis of the parameters $t_{20}$, $t_{15}$, and $t_{10}$, it can be seen that the values for SC 10-15 and 24, except for SC 11, are significantly greater than those for SC 16-23, except the SC 16 and 17 for $t_{10}$ (see Figure 8). This fact might suggest a modal behavior such as it occur with the parameters $t_3$ and $t_4$. On the other hand, the linear fits between these parameters and *RM* and *Rm* suggest that high values of $t_{10}$, $t_{15}$, and $t_{20}$ seem to be associated with low values of *RM* and *Rm* and *vice versa* (see Figure 9). Moreover, it can be seen that all data with $t_{20}$ values greater than the median (4 months) have *Rm* values less than its median (15.3), and all data with $t_{20}$ values greater than 4 months (the $t_{20}$ median) have *Rm* values greater than 15.3 (the *Rm* median). The most significant correlation found is for the linear fits between *Rm* and $t_{10}$ ($r = -0.85$; *p*-value $< 0.001$) but we want to highlight although the correlation is strong, the estimated values are sometimes far from the observed values. As it occurs with the parameters $t_{10}$, $t_{15}$, and $t_{20}$, the behavior of SC 24 for the values of the parameters $t_{20f-l}$, $t_{15f-l}$, and $t_{10f-l}$ differs with respect to the previous six or seven solar cycles since these parameters for SC 24 increase significantly their values (see Figure 10). Again, it might be explained with a change in a modal behaviour. On the other hand, it seems to be a small cyclical behaviour approximately equal to four cycles for $t_{15l\text{-}M}$ and $t_{10l\text{-}M}$, but not so clearly for $t_{20l\text{-}M}$ (see Figure 11). Finally, we obtained data more normally distributed for the fits between $t_{10l\text{-}M}$, $t_{15l\text{-}M}$, and $t_{20l\text{-}M}$ and $t_{10f\text{-}l}$, $t_{15f\text{-}l}$, and $t_{20f\text{-}l}$, and a more significant correlation for $RM$ against $t_{10f\text{-}l}$, $t_{15f\text{-}l}$, and $t_{20f\text{-}l}$ (see Figure 12). Thus, high $RM$ values seem to be associated with small $t_{10f\text{-}l}$, $t_{15f\text{-}l}$, and $t_{20f\text{-}l}$ values but we want to highlight that estimates for future solar cycles from these parameters may not be accurate.

Therefore, we conclude, on the one hand, that the spotless days could be a good approach to understanding the behavior of some parameters of the solar activity. However, they should be used with caution for the prediction of solar activity because the relative error in some of the predictions was found to be large. On the other hand, we have obtained a possible mode change in the SC 23, a break in the behavior with respect to the six or seven previous solar cycles and more similar to the first solar cycles considered in this study (SC 10-15). We want to highlight that other papers have also noted other peculiarities for that recent solar minimum (Cranmer, Hoeksema, and Kohl, 2010).

**Acknowledgements**

This research was supported by the Economy and Infrastructure Counselling of the Junta of Extremadura through project IB16127 and grant GR15137 (co-financed by the European Regional Development Fund) and by Ministerio de Economía y Competitividad of the Spanish Government (AYA2014-57556-P).

**Disclosure of Potential Conflicts of Interest**

The authors declare that they have no conflicts of interest.